\documentstyle[12pt,epsf,epsfig]{article}
\def\be{\begin{equation}}
\def\ee{\end{equation}}
\def\ba{\begin{eqnarray}}
\def\ea{\end{eqnarray}}
\def\la{~\mbox{\raisebox{-.6ex}{$\stackrel{<}{\sim}$}}~}
\def\ga{~\mbox{\raisebox{-.6ex}{$\stackrel{>}{\sim}$}}~}
\def\bq{\begin{quote}}
\def\eq{\end{quote}}

 at 10truept

\newcommand{\beq}{\begin{equation}}
\newcommand{\eeq}{\end{equation}}
\newcommand{\beqa}{\begin{eqnarray}}
\newcommand{\eeqa}{\end{eqnarray}}

\def\la{~\mbox{\raisebox{-.6ex}{$\stackrel{<}{\sim}$}}~}
\def\ga{~\mbox{\raisebox{-.6ex}{$\stackrel{>}{\sim}$}}~}

\def\ltap{\ \raise.3ex\hbox{$<$\kern-.75em\lower1ex\hbox{$\sim$}}\ }
\def\gtap{\ \raise.3ex\hbox{$>$\kern-.75em\lower1ex\hbox{$\sim$}}\ }
\def\gl{\ \raise.5ex\hbox{$>$}\kern-.8em\lower.5ex\hbox{$<$}\ }
\def\roughly#1{\raise.3ex\hbox{$#1$\kern-.75em\lower1ex\hbox{$\sim$}}}

\begin{document}
\thispagestyle{empty}
\begin{flushright}
hep-th/0307016\\ June 2003
\end{flushright}
\vspace*{1cm}
\begin{center}
{\Large \bf Primeval Corrections to the CMB Anisotropies}\\
\vspace*{1.2cm} {\large Nemanja
Kaloper\footnote{kaloper@physics.ucdavis.edu}
and Manoj Kaplinghat\footnote{kaplinghat@ucdavis.edu} }\\

\vspace{.8cm}
{\em Department of Physics, University of California}\\
{\em Davis, CA 95616}\\
\vspace{1.5cm} ABSTRACT
\end{center}
We show that deviations of the quantum state of the inflaton from
the thermal vacuum of inflation may leave an imprint in the CMB
anisotropies. The quantum dynamics of the inflaton in such a state
produces corrections to the inflationary fluctuations, which may
be observable. Because these effects originate from IR physics
below the Planck scale, they will dominate over any
trans-Planckian imprints in any theory which obeys decoupling.
Inflation sweeps away these initial deviations and forces its
quantum state closer to the thermal vacuum. We view this as the
quantum version of the cosmic no-hair theorem. Such imprints in
the CMB may be a useful, independent test of the duration of
inflation, or of significant features in the inflaton potential
about $60$ e-folds before inflation ended, instead of an unlikely
discovery of the signatures of quantum gravity. The absence of any
such substructure would suggest that inflation lasted
uninterrupted much longer than ${\cal O}(100)$ e-folds.

\vfill \setcounter{page}{0} \setcounter{footnote}{0}
\newpage

Generic models of inflation produce a lot of accelerated
expansion. They far surpass the minimum needed to solve the
horizon and flatness problems, of the order of $N \simeq 60$. For
example in the case of chaotic inflation driven by a power-law
potential $\lambda \phi^n/n$, one finds that typically $N \ga
10^{24/n} \gg {\cal O}(100)$. This yields perhaps the most robust
of all inflationary predictions, that the universe should be
spatially flat, with $\Omega _{\rm matter} + \Omega_{\rm DE} =
1$~\cite{infl}. More generally, this is usually taken to mean that
inflation acts as a powerful amnesia tonic, efficiently relieving
the universe of the memory of its initial state. Having fewer
e-folds, or producing significant changes in the inflaton sector
mid-way through inflation, requires fine-tunings of the initial
conditions and/or the inflaton sector beyond those which are
deemed acceptable by the current lore \cite{clos}. A more
optimistic stance could be that any signs of short inflation or
new dynamics during it is an indication of some as yet unknown new
physics, which the inflaton is sensitive to. Although such
phenomena may appear like fine-tuning by the current lore, one
could hope to identify the underlying physics with better
understanding of inflation. Hence either short inflation or
changes of the inflaton dynamics $\sim 60$ e-folds before the exit
at this point cannot be taken as a robust
prediction\footnote{Possibly excepting anthropic arguments, about
which we are agnostic at the moment.} but as an indication of
something special about inflationary dynamics and/or the initial
conditions.

On the other hand, we can take a bottom-up approach to conceptual
cosmology, and simply ask if we can measure for how long the final
stage of inflation went on uninterrupted. For example, the current
cosmological observations are indirectly sensitive to short
inflation, because it could leave nonvanishing spatial curvature
of the universe. At present the observations limit the spatial
curvature to be at most a few percent of the total $\Omega$, and
the bounds are a little bit weaker if the curvature is positive
\cite{boomer,maxima,wmap}. The bounds will be improved some in the
future \cite{eht}. Thus it would be interesting to consider
alternative probes of the length of inflation, or of significant
features in the inflaton dynamics.

In this note we show that substructure in the CMB anisotropies
could provide us with another probe of inflation some $\sim 60$
e-folds before the end. If inflation was interrupted $\sim 60$
e-folds before the exit by environmental conditions, induced
either by a non-inflationary stage, or by a change of slow roll
parameters, the quantum state of the inflaton during the
generation of the inflationary fluctuations was not the usual
thermal vacuum, but included some deviations from it. These
effects may resemble classical inhomogeneities, in that they can
be viewed as lumps of energy on top of the ground state, and can
be represented as coherent state excitations of the thermal
vacuum. They may also be intrinsically quantum, encoding initial
phase correlations arranged by quantum effects before inflation,
or by the dynamics which may have intervened at the onset of the
last 60 e-folds. The latter effects can be represented as squeezed
states, which have been prepared by primordial quantum effects
preceding inflation\footnote{We thank J.D. Bjorken for a very
useful discussion of this issue.}.

We will demonstrate explicitly how such a squeezed state arises
from the kinks in the inflaton slow roll parameters. During the
onset of the final stage of inflation the frequencies of the
inflaton eigenmodes change in time slightly non-adiabatically.
This induces a Bogoliubov transformation between the modes before
and after the transition, and therefore between their
corresponding annihilation and creation operators. We will take
the initial inflaton state to be the ground state of the theory
just before the transition, because the transition takes only
${\cal O}(1)$ Hubble times to complete. Therefore it is basically
a sudden transition for modes with horizon-size wavelengths, so
that the system remains in the state it occupied before the
transition. Because of the Bogoliubov transformation, this state
is a squeezed state on top of the thermal vacuum defined by the
theory after the onset of the final stage of inflation.

The intuitive picture of this dynamics is akin to the
quantum-mechanical system in a deep potential well, whose depth is
suddenly increased. Prior to the change the system settles in its
ground state, given by the minimum energy state in the well.
Because the transition is fast, the system remains trapped in this
state. However, after the transition, this state will not be the
minimum energy state any more, since the depth of the well
increased. Hence the system will be in an excited state on top of
the new ground state. When the inflaton is quantized in such a
state, its deviations from the thermal vacuum will correct the
standard thermal vacuum result. They may be stronger than the
imprints from new physics computable by effective field theory
\cite{kkls}. We find that such effects contribute a factor
$\Bigr(1 + \Delta (\eta_H - \epsilon_H) (H/p) \sin(2p/H) + \ldots
\Bigl)$ to the thermal vacuum result, where $\ldots$ stand for
additional slow roll and adiabatic corrections. Here $H$ is the
Hubble scale during inflation, $p \ga H$ is the physical momentum
of the fluctuation at the moment of transition, $\epsilon_H =
{\dot \phi^2}/[{2 m^2_P H^2}]$ and $\eta_H = - \ddot \phi/[H\dot
\phi]$ are the slow roll parameters, $\Delta (\eta_H -
\epsilon_H)$ is the change of their difference at the transition
between two stages, $\phi$ is the inflaton vev, overdot is a
derivative with respect to the comoving time $t$, and $m^2_P =
8\pi/G_N$ is the reduced Planck mass. We note that in long
inflation with very kinky structure in the inflaton potential a
large change in $\epsilon_H$ would have even more dramatic
consequences already in the leading order density contrast
\cite{star92,adams01}, rendering the subleading corrections
irrelevant. This can be seen by rewriting $\delta \rho/\rho$ as
$\delta \rho/\rho \propto H/[\sqrt{\epsilon_H} m_P]$, and noting
that it would change a lot if $\epsilon_H$ jumped while $H$ stayed
fixed. Since we are interested in the subleading corrections on
top of the standard result, we will ignore long inflation which
had such a strong jump in $\epsilon_H$ mid-course. However our
treatment gives a way of distinguishing the models with a milder
variation of $\epsilon_H$, but a large change in $\eta_H$, that
would not affect so strongly the leading order result. We will
also comment on the possibility of softer features in the inflaton
potential which could mimic the effects we consider already at the
leading order.

Our signal might remind one of the effects recently claimed to
arise from the trans-Planckian physics during inflation (for
various approaches and discussions see, e.g.
\cite{bramart}-\cite{golo} and references therein, and for other
ways to get similar signals see \cite{burgess,cr}). A simple
framework for the formulation of such effects is provided by the
$\alpha$-vacua \cite{alpha,danielsson,golo}. However the
short-distance behavior of quantum field theory in $\alpha$-vacua
forces one to abandon locality and decoupling in order to regulate
the theory in the UV, once interactions are included
\cite{baman}-\cite{hol}. This means that the theory cannot be kept
under full calculational control as an interacting quantum field
theory\footnote{The cosmological constant problem is, in our
opinion, still too much of a mystery to be taken as a clear-cut
directive for abandoning decoupling  in hope that this would
simultaneously fix all the problems with $\alpha$-vacua.},
conflicting with the usual notion of decoupling. Further problems
arise from considering the diffuse gamma-ray background measured
by EGRET which already excludes the possibility of detectable
imprints of $\alpha$-vacua in the CMB \cite{star}. A proposal for
avoiding this was offered in \cite{danielsson}, but it mandates
changing $\alpha$, and so the vacuum, in time as a function of the
dominant source of energy density in the Universe, which is again in
conflict with decoupling. Thus it is difficult
to regard the results of a naive perturbation theory around
$\alpha$-vacua \cite{danielsson,golo} as predictions for
signatures of quantum gravity.

Our result differs in crucial ways. We obtain it by computing the
fluctuations in an excited state on top of the usual thermal
vacuum, generated by the intervening low energy evolution. The
momentum $p$ in it is not some fixed trans-Planckian scale, but
the physical momentum of the fluctuations expelled out of the horizon,
evaluated at the transition, and so the imprint decreases with
their wavelength. Further there is the $\Delta(\eta_H-\epsilon_H)$
suppression. Thus our imprints originate completely from the IR
physics below the Planck scale. As long as quantum gravity yields
the usual effective quantum field theory below the Planck scale
obeying decoupling, which we assume here, these IR effects provide
the dominant influence on the CMB, irrespective of the details of
short distance dynamics. Inflation pushes the universe into the
thermal vacuum; the longer the inflationary stage, the closer to
the thermal vacuum were the state during which the observable
fluctuations in the CMB background were produced. This is the
quantum-mechanical version of the {\it cosmic no-hair} theorem. It
implies that the signatures of quantum deviations from the thermal
vacuum could be very sensitive to the duration of inflation. Our
results provide a very simple, yet general, demonstration of this,
complementing \cite{burgess}.

The main implication of our analysis for observations is that if
for any reason $\eta_H - \epsilon_H$ changed significantly $\sim
60$ e-folds before the end of inflation, the effects of such a
change in the CMB may be visible in the horizon-scale fluctuations
today. If inflation were short, and the universe had a spatial
curvature close to observability, with $\Omega_{\rm k}$ of few
percent, the effects we consider should be within observer's
reach. We specifically note that such effects may lead to the
reduction of power in the low $\ell$ CMB multipoles in some
models of inflation. Conversely, the absence of any such
substructure would strongly suggest that inflation lasted
uninterrupted much more than the minimum 60 e-folds. A long
inflation would wipe out any spatial curvature and produce
$\Omega_{\rm matter} + \Omega_{\rm DE} = 1$. The memory of the
quantum correlations in the initial state of the universe would be
deeply buried in very small effects which would be extremely
efficiently obscured by post-inflationary nonlinear dynamics.

We now turn to our setup. We will use a very simple toy model,
where we imagine that the universe can be described for the most
part by a (spatially flat) FRW line element, perturbed by initial
inhomogeneities and later by inflationary fluctuations. We confine
our analysis to the regime where the perturbations are small,
${\delta \rho}/{\rho} \ll 1$. We further imagine that some agent
altered inflationary dynamics $\sim 60$ e-folds before the end.
Either inflation was short, following an epoch of decelerated
expansion, such as e.g. a brief radiation era after the primordial
singularity, or the inflaton went over a potential bump which
changed the slow roll parameters. The differences in the quantum
dynamics of the inflaton before and after this transition are
captured by the time-variation of the frequencies of the inflaton
eigenmodes. Before and after inflation, their form is controlled
by different backgrounds. The rapid change of the background
environment during the transition will induce a slightly
non-adiabatic contribution to the frequencies. Thus the eigenmodes
will also be modified non-adiabatically. This will change the
Hilbert space\footnote{Here we will not dwell on the conceptual
problems involved in defining the Hilbert space of the inflaton in
the first place \cite{tw,hks,witt,lenny}.} of the theory by
inducing a Bogoliubov rotation of the annihilation and creation
operators. The quick onset of the last stage of inflation enables
us to treat the transition at the quantum level as basically a
sudden transition, where the quantum system remains in the state
which it occupied just before the transition. If we take the
quantum state of the inflaton just before the transition to be the
vacuum of the theory then, this state will be a squeezed state on
top of the thermal vacuum after the transition. Therefore the
inflaton state from which the inflationary fluctuations originate
is populated with long-wavelength quanta of the inflaton with
quantum correlations prearranged by the preceding evolution. This
is of course an idealized choice; in reality the inflaton state at
this instant may be an excitation of the vacuum, which may contain
inhomogeneties etc. Such a general state can be viewed as a
coherent state on top of our initial vacuum. Since our purpose
here is to explore the quantum correlations in the initial vacuum
itself, we will ignore the excitations of this state because they
behave like localized inhomogeneities in the inflating patch,
which will get inflated away as usual. This simplified analysis is
sufficient to illustrate our main points. We expect that a more
detailed analysis will share the qualitative features of our main
results as long as inflation is short.

The effective field theory description of the quantum fluctuations
of the inflaton, which we rely on, will break down at the cutoff,
say at the string scale, but because of decoupling this does not
produce significant effects on the horizon scales, where the
initial correlations dominate. This is reflected in our
calculation by the fact that the correlations are suppressed by a
power of the momentum, and so drop off at short distances. This in
turn means that the initial state is regulated in the UV in the
usual way, and is not subject to the maladies plaguing
$\alpha$-vacua discussed in \cite{baman}-\cite{hol}. In the limit
of eternal de Sitter space, these correlations would completely
disappear and the initial state becomes precisely the usual
thermal vacuum, instead of one of the $\alpha$-vacua. A
consequence of this is that as inflation proceeds, shorter
wavelength modes are expelled out of the apparent horizon. These
modes encode progressively less information about the initial
quantum deviations. Hence the state of the inflationary universe
when these fluctuations are frozen appears closer to the thermal
vacuum. We view this as a quantum-mechanical version of the
cosmic no-hair theorem.

We first briefly review the gauge-invariant perturbation theory
for inflation \cite{bard}-\cite{mfb}. Much of the useful formalism
of the evolution of fluctuations is also exhibited in
\cite{grishchuk,andy,polsta}. We restrict our attention to the
scalar perturbations in longitudinal gauge which is sufficient for
our purposes. Once the results are expressed in terms of the
gauge-invariant variables, one can change the gauge at will
anyway. Thus the line element is \be ds^2 = a^2(\eta) \Bigl[
-\Bigl(1+2{ \Psi }(\eta,\vec x)\Bigr) d\eta^2 +
\Bigl(1+2{\Phi}(\eta,\vec x)\Bigr) d\vec x^2 \Bigr] \, .
\label{flucfrw} \ee The conformal time $\eta$ is related to the
usual comoving FRW time $t$ by $dt = a d\eta$. In the longitudinal
gauge, the two metric perturbations $\Psi,\Phi$ coincide with
gauge-invariant potentials for the perturbations. In general, they
are not independent. However, their detailed relationship depends
on the matter contents of the universe that sets the background of
(\ref{flucfrw}).

During inflation, the dominant source of the stress-energy in the
Einstein's equations is the inflaton field. The inflaton field
sector can be written as
\be
\phi(\eta,\vec x) = \phi(\eta) + \delta \phi(\eta,\vec x) \, .
\label{inflfluc}
\ee
The independent
background equations are, using conformal time variables,
\ba
&& 3m^2_P{\cal H}^2 = \frac{(\phi')^2}{2}
+ a^2 V(\phi) \, , \nonumber \\
&& \phi'' + 2 {\cal H}\phi' + a^2 \frac{\partial V}{\partial
\phi} = 0 \, ,
\label{back}
\ea
where the prime denotes the
derivative with respect to the conformal time and $ {\cal H}=
a'/a$. A detailed analysis of the perturbations yields $\Psi = -
\Phi$. Moreover, the potential $\Phi$ and the inflaton
perturbation $\delta \phi$ are related by momentum
conservation as
\be
\phi' \delta \phi = - 2 m^2_P (\Phi' + {\cal H}\Phi) \, .
\label{fluctrels}
\ee
Hence during inflation only one of the perturbations $\Phi, \delta
\phi$ is independent. One chooses it such that it has the
canonical commutation relations. The quantum mechanical
calculation then links $\Phi$ to the properties of the inflaton
effective action and the quantum state of the inflaton during the
period of inflation when the fluctuations are produced
\cite{mukh,perts}.

To determine the effect of the perturbations in an all-inclusive
way, summing all the contributions to the ripples in the spacetime
caused by the inflaton fluctuations, one performs an infinitesimal
diffeomorphism $\delta \eta = \delta \rho/\rho' = \delta
\phi/\phi'$ \cite{diffeo}, because during inflation $\rho =
V(\phi)$, $\delta \rho = \partial_\phi V \delta \phi$ and $\rho'
= \partial_\phi V \phi'$ to the leading order in the slow roll
parameters. In this new gauge, the curvature perturbation is the total
perturbation of the $\eta = {\rm const}$ hypersurfaces, and it is
given in terms of another gauge-invariant potential $\Theta = \Phi
- \frac{\cal H}{\phi'} \delta \phi $ as
\be
\frac{\delta {\cal R}_3}{R} = - \frac{1}{3a^2 H^2} \vec \nabla^2
\Theta(\eta,\vec x) \, ,
\label{totpert}
\ee
where $H = \dot a/a = {\cal H}/a$ is the
comoving Hubble parameter and $\vec \nabla$ denotes
derivatives with respect to the spatial coordinates $\vec x$. The
canonically normalized scalar field corresponding to this
perturbation, which is to be promoted into the quantum
inflaton field, is
\be
\varphi = a \delta \phi - \frac{a \phi'} {\cal H} \, \Phi =
- {\cal Z} \Theta  \, ,
\label{scalarinfl}
\ee
which is clearly gauge-invariant, being defined in terms of
$\Theta $. Following the common practice we have defined $ {\cal
Z} = \frac{a \phi'} {\cal H}= \frac{a \dot \phi} {H}$
\cite{mukh,mfb}. Because the unperturbed background is spatially
flat, we can expand all the fields in Fourier modes, $ f_{\vec
k}(\eta) = \int \frac{d^3\vec x}{(2\pi)^{3/2}} f(\eta, \vec x)
e^{-i \vec k \cdot \vec x} $. Then $\frac{\delta {\cal R}_3}{R}(k)
= \frac{k^2}{3a^2 H^2} \Theta_{\vec k}(\eta) $, and the
definition of the power spectrum ${\cal P}(k)\delta^{(3)}(\vec k -
\vec q) = \frac{k^3}{2\pi^2} \langle \Theta_{\vec k}(\eta) \,
\Theta^\dagger{}_{\vec q}(\eta) \rangle $ gives
\be
{\cal P}(k)\delta^{3}(\vec k - \vec q) = \frac{k^3}{2\pi^2}
\Bigl(\frac{H}{\dot \phi}\Bigr)^2 \langle \frac{\varphi_{\vec k}}{a} \,
\frac{\varphi^\dagger{}_{\vec q}}{a} \rangle \, ,
\label{power}
\ee where $\langle {\cal O} \rangle$ stands for the
quantum expectation value of the 2-point operator ${\cal O}$ in
the quantum state of inflation.

The scalar field (\ref{scalarinfl}) is the properly defined,
gauge-invariant small fluctuation of the inflaton. In perturbation
theory its dynamics is governed by the quadratic action
\be
S_{\varphi} = \frac12 \int d\eta d^3 \vec x \, \Bigl( (\varphi')^2 -
 (\vec \nabla \varphi)^2 + \frac{{\cal Z}''}{\cal Z}
{\varphi^2} \Bigr) \, .
\label{action}
\ee
To quantize the theory, we use the field and its conjugate
momentum in the momentum picture:
\ba
\varphi_{\vec k}(\eta) &=& \int \frac{d^3 \vec x}{(2\pi)^{3/2}} \,
\varphi(\eta, \vec x) \, e^{-i \vec k \cdot \vec x} \, , \nonumber \\
\pi_{\vec k}(\eta) &=& \int \frac{d^3 \vec x}{(2\pi)^{3/2}} \,
\pi(\eta, \vec x) \, e^{-i \vec k \cdot \vec x} \, .
\label{defmomenta}
\ea
Note that $\varphi^\dagger_{\vec k} = \varphi_{-\vec k}$,
$\pi^\dagger_{\vec k} = \pi_{-\vec k}$. From (\ref{action})
we have $\pi = \varphi'$, which using
(\ref{defmomenta}) translates to $\pi_{\vec k}(\eta) =
\varphi'_{\vec k}(\eta)$ for the Fourier transforms.
One can check that the
canonical commutation relations $[ \varphi(\eta, \vec x),
\pi(\eta, \vec y) ] = i \delta^{(3)}(\vec x - \vec y)$ imply
$[ \varphi_{\vec k}(\eta), \pi^\dagger_{\vec q}(\eta) ]
= i \delta^{(3)}(\vec k - \vec q)$.
The Hamiltonian is
\be
H_{\varphi} = \frac12 \int d^3 \vec k \, \Bigl(  {\pi^\dagger_{\vec k}} \,
\pi_{\vec k} +
\bigl( k^2 - \frac{{\cal Z}''}{\cal Z} \bigr) \,
{\varphi^\dagger_{\vec k}} \, \varphi_{\vec k} \Bigr) \, .
\label{phiham}
\ee
The Fourier modes of $\varphi$ obey the field
equation\footnote{It is easy to
find the asymptotic behavior of the solutions of (\ref{scalfe}) in
the general case. To the leading order, in the short wavelength
limit, $k^2 \gg \frac{{\cal Z}''}{\cal Z}$, one finds $\varphi_k
\rightarrow A_k \cos(k \eta + \theta_k)$, while in the limit of
long wavelengths, where $k^2  \ll \frac{{\cal Z}''}{\cal Z}$, the
result is $\varphi_{\vec k} \rightarrow {B_k}{\cal Z} + C_k {\cal
Z} \int \frac {d \eta}{{\cal Z}^2}$. From (\ref{scalarinfl}) the
curvature perturbation is \cite{mukh}
$$
\Theta_{\vec k} =  - \cases{ \Bigl(\frac{H}{\dot \phi}\Bigr)
\frac{A_k}{a} \cos(k \eta + \theta_k) & if $k^2 \eta^2 \gg 2 $ \cr
& \cr B_k + C_k \int \frac {d \eta}{{\cal Z}^2}~~ & if $k^2 \eta^2
\ll 2$} \, \,
$$
The $\propto C_k$ mode in the latter case is the decaying
superhorizon mode. From this it is clear than any short
wavelength, oscillatory mode excited inside the apparent horizon
$H^{-1}$ will end up expelled out of it by inflationary
stretching, where it will freeze out retaining a nearly constant
amplitude $B_k$, producing a nearly scale invariant spectrum of
perturbations \cite{mukh,perts}.
Thermodynamics of this process of modes leaking out of the
apparent horizon during inflation, which resembles a leaky can,
has been discussed in \cite{aks}.} \cite{mukh,mfb} (where $k =
|\vec k|$)
\be
\varphi_{\vec k}'' + \bigl(k^2 - \frac{{\cal Z}''}{\cal Z} \bigr)
\, \varphi_{\vec k} = 0 \, .
\label{scalfe}
\ee
The mode expansion of the field $\varphi(\eta,\vec x)$ is
\be
\varphi(\eta,\vec x) = \int \frac{d^3 \vec k}{(2\pi)^{3/2}}
\Bigl(b(\vec k) u_k(\eta) e^{i\vec k \cdot \vec x}
+ b^\dagger(\vec k) u^*_k(\eta) e^{-i\vec k \cdot \vec x} \Bigr) \, ,
\label{scalarinflq}
\ee
where the annihilation and creation operators
$b(\vec k), b^\dagger(\vec q)$ satisfy the usual operator algebra
\be
[ \, b(\vec k), \, b^\dagger(\vec q)\, ] =
\delta^{(3)}(\vec k - \vec q) \, ,
~~~~~~~~~~ [\, b(\vec k), \, b(\vec q)\, ] =
[\, b^\dagger(\vec k), \, b^\dagger(\vec q)\, ] = 0 \, .
\label{alg}
\ee
The orthogonal eigenmodes $u_k, u^*_k$ of (\ref{scalfe}) are
easy to construct in the slow roll approximation, when
$\epsilon_H = \dot{\phi}^2/{2 m^2_P H^2}$ and $\eta_H = -\ddot{\phi}/H
\dot{\phi}$ are small, $\epsilon_H, \eta_H \ll 1$. During
slow roll inflation, to ${\cal O}(\epsilon_H, \eta_H)$,
we have $(1-\epsilon_H)\eta = -
\frac{1}{aH}$ and so $\frac{{\cal Z}''}{\cal Z} = \frac{2 -
3 \eta_H + 6 \epsilon_H }{\eta^2} $.
Then the mode equation (\ref{scalfe}) becomes \cite{lyste}
\be
u_k'' + \bigl(k^2 - \frac{2 - 3 \eta_H +
6 \epsilon_H }{\eta^2} \bigr) \, u_k = 0 \, .
\label{eigen}
\ee
The standard choice \cite{mukh,perts}
of the eigenmodes $u_k, u_k^*$ is to take
\ba
u_k (\eta) &=& - \, \frac{\sqrt{\pi \eta}}{2} \,
H^{(-)}_{\nu}(k \eta) \, , \nonumber \\
u_k^* (\eta) &=& - \, \frac{\sqrt{\pi \eta}}{2} \,
H^{(+)}_{\nu}(k \eta)  \, ,
\label{eigensols}
\ea
as the positive and negative frequency modes, respectively, where
$\nu = 3/2 - \eta_H + 2\epsilon_H$ \cite{lyste}. The normalization
of $u_k$ is chosen such that Eq. (\ref{alg}) follows from the
canonical commutation relations $[\, \varphi(\eta,{\vec x}), \,
\pi(\eta,{\vec y}) \, ] = i\delta^3(\vec x - \vec {y})$.
Substituting $u_k, u_k^*$ in (\ref{scalarinflq}) amounts to
choosing the thermal vacuum $|0 \rangle $ as the ground state of
the theory, because it is annihilated by the operators
corresponding to the positive frequency modes in
(\ref{eigensols}):
\be
b(\vec q) | 0 \rangle = 0 \, .
\label{vac}
\ee
Using $| 0 \rangle$ as the state of the inflaton during
inflation and ignoring the slow roll corrections, in which
case the eigenmodes (\ref{eigensols}) reduce to
\ba
u_k &=& \frac{1}{\sqrt{2k}}\Bigl(1 - \frac{i}{k\eta} \Bigr) \,
e^{-ik\eta} \, , \nonumber \\
u_k^* &=& \frac{1}{\sqrt{2k}} \Bigl(1 + \frac{i}{k\eta} \Bigr) \,
e^{ik\eta} \, ,
\label{modes}
\ea
leads to $ \langle 0 | \frac{\varphi_{\vec k}}{a}
\frac{\varphi^\dagger{}_{\vec q}}{a} | 0 \rangle = \frac{2\pi^2}{k^3}
\bigl(\frac{H}{2\pi}\bigr)^2 \delta^{(3)}(\vec k - \vec q)$,
yielding the standard result for the power spectrum,
\be
{\cal P}(k) = \Bigl(\frac{H}{\dot \phi}\Bigr)^2
\Bigl(\frac{H}{2 \pi}\Bigr)^2 \, .
\label{powerst}
\ee
The corrections from slow roll effects, new physics and
initial quantum correlations come in the form of multiplicative
factors which can be calculated within a given
theory.

In what follows we will focus on the corrections from quantum
correlations and slow roll effects. In short inflation the
evolution did not have enough time to prepare the system in the
thermal vacuum. One could instead take the instantaneous vacuum
defined by the Heisenberg operators $b(\vec k, \eta_0),
b^\dagger(\vec k, \eta_0)$ at the time when inflation began
\cite{mfb}. This would reproduce the thermal vacuum in the limit
$\eta_0 \rightarrow -\infty$. However as we have explained above,
there are additional effects coming from the background evolution.
During the transition the function $\frac{{\cal Z}''}{\cal Z}$ in
the mode equation (\ref{scalfe}) changes rapidly, inducing
modifications of the frequencies of the inflaton eigenmodes. This
results in a Bogoliubov transformation in the Hilbert space which
also includes non-adiabatic contributions. The net effect is that
because the transition is quick the inflaton is trapped in the
state it occupied just before the onset of the last stage of
inflation, which is now a squeezed state on top of the inflaton
thermal vacuum because of the transition-induced Bogoliubov
transformation. Below we will compute the corrections from the
quantum correlations in this state. We will also include the slow
roll corrections, computed in \cite{lyste}, and the corrections
from the adiabatic evolution of the vacuum, because they may be
numerically significant. We will ignore possible contributions
from the initial inhomogeneities, because while they will
typically be important at shorter scales, they are quickly
inflated away. Thus they should not affect the fluctuations at
horizon crossing after a few e-folds.

Let us now define the initial state of the inflaton. As we said
above, we take the initial state of the last $\sim 60$ e-folds of
inflation to be the instantaneous vacuum just before the onset of
this stage. This state is annihilated by the Heisenberg picture
annihilation operator at the time $\eta_0^-$, which denotes the
instant just before the transition. The Heisenberg picture
annihilation and creation operators obey the canonical commutation
relations $[ b(\vec k, \eta), \, b^\dagger(\vec q, \eta)] =
\delta^{(3)}(\vec k - \vec q)$, $[ b(\vec k, \eta), \, b(\vec q,
\eta)] = [ b^\dagger(\vec k, \eta), \, b^\dagger(\vec q, \eta)] =
0$, and can be defined in terms of the fields and their conjugate
momenta. They are \cite{andy,polsta}
\ba
b(\vec k, \eta) &=& \frac1{\sqrt{2}}
\Bigl( \sqrt{k} \, \varphi_{\vec k}(\eta)
+ \frac{i}{\sqrt{k}} \, \pi_{\vec k}(\eta) \Bigr) \, ,
\nonumber \\
b^\dagger(\vec k, \eta) &=& \frac1{\sqrt{2}}
\Bigl( \sqrt{k} \,
\varphi^\dagger_{\vec k}(\eta) - \frac{i}{\sqrt{k}} \,
\pi^\dagger_{\vec k}(\eta) \Bigr) \, .
\label{ancreops}
\ea
Thus the instantaneous vacuum $| \tilde 0 \rangle$
of the theory just before
the last stage of inflation obeys
\be
b(\vec k, \eta^-_0) | \tilde 0 \rangle = 0 \, .
\label{defvac}
\ee
We note here that we are interested only in the leading order
contributions from the non-thermal effects induced by the sharp
transition. We will see that
the slow roll corrections, the contributions from
initial quantum correlations and the corrections from adiabatic
evolution come with their own small parameters: powers of
$\epsilon_H, \eta_H$; $\Delta(\eta_H - \epsilon_H) H/p \,$; and
$(H/p)^2$ respectively. Therefore in computing the leading order
form of each one we can ignore the others. This allows us to use
the massless eigenmodes in place of the exact Hankel functions in
(\ref{eigensols}) when computing the corrections from adiabatic
evolution and initial correlations, and only use the slow
roll-improved eigenmodes (\ref{eigensols}) when considering the
slow roll corrections. We will include these slow roll corrections
because they may be numerically significant by simply incorporating
the known result for the slow roll corrections from
\cite{lyste}. A more general calculation accounting for
interference terms may be interesting in order to get a more
suitable framework for data fits, but is beyond the scope of the
present work.

To determine the effects of the transition on the Hilbert space,
and specifically on the state $| \tilde 0 \rangle$ (\ref{defvac}),
we consider the field equation (\ref{scalfe}) in a general
environment. One can show that in general
\be
\frac{{\cal Z}''}{\cal Z} =
\frac{a''}{a} +
\frac12 \bigl(\frac{\epsilon'_H}{\epsilon_H}\bigr)'
+ \frac{\epsilon'_H}{\epsilon_H} {\cal H}  +
\frac14 \bigl(\frac{\epsilon'_H}{\epsilon_H}\bigr)^2 \, .
\label{zs}
\ee
To find the effects of the sudden transition on the Hilbert space,
we need to evolve the Heisenberg operators of the theory through
the transition using the field equation (\ref{scalfe}) with the
general form of $\frac{{\cal Z}''}{\cal Z}$ {\it included}.
Indeed, from inspecting (\ref{zs}), it is clear that the
contribution to $\frac{{\cal Z}''}{\cal Z}$ coming from $\frac12
\bigl(\frac{\epsilon'_H}{\epsilon_H}\bigr)'$ experiences a jump!
Because $\frac{\epsilon'_H}{\epsilon_H}$ is linear in $\epsilon_H,
\eta_H$, the difference of $\frac{\epsilon'_H}{\epsilon_H}$ before
and after the transition will be of the same order as the quantity
itself. This is the source of the non-adiabatic evolution of the
operators. Basically, the environment pumps some energy into them
during the transition. To determine the transformation, we can
treat the problem as a Schr\"odinger problem with a piecewise
smooth potential:
\be
\varphi_{\vec k}'' +
\Bigl(k^2 - V(\eta) \Bigr) \, \varphi_{\vec k} = 0 \, ,
\label{schro}
\ee
where
\be
V(\eta) = \frac{a''}{a} +
\frac12 \bigl(\frac{\epsilon'_H}{\epsilon_H}\bigr)'
+ \frac{\epsilon'_H}{\epsilon_H} {\cal H}  +
\frac14 \bigl(\frac{\epsilon'_H}{\epsilon_H}\bigr)^2 \, .
\label{potential}
\ee
The problem of matching the operators is completely analogous to
the quantum mechanical problem of a particle scattering on a
potential bump. Rather than looking at the specifics of an explicit
construction of the modes for any given environment before the
last $\sim 60$ e-folds of inflation, we take the shortcut and find
the effect of the transition directly on the field operators. To
do this, we impose the continuity of $\varphi_{\vec k}$ across the
transition in the usual way, and determine the jump of
$\varphi'_{\vec k}$ as dictated by $\frac12
\bigl(\frac{\epsilon'_H}{\epsilon_H}\bigr)'$ in the  Gaussian
pillbox integration of (\ref{schro}). Denoting the quantities
slightly before the transition by the argument $\eta_0^-$ and
those slightly after by the argument $\eta_0^+$, the integration
yields
\be
\varphi'_{\vec k}(\eta^+_0) = \varphi'_{\vec k}(\eta^-_0)
+ \frac{{\epsilon^+_H}'}{2\epsilon^+_H} \, \varphi_{\vec k}(\eta_0)
- \frac{{\epsilon^-_H}'}{2\epsilon^-_H} \, \varphi_{\vec k}(\eta_0)
\, .
\label{intbcs}
\ee
The slow roll parameter $\epsilon_H$ obeys
$\frac{\epsilon'_H}{\epsilon_H} = - 2{\cal H} (\eta_H - \epsilon_H)$.
This gives the jump conditions
\ba
\varphi_{\vec k}(\eta^+_0)
&=& \varphi_{\vec k}(\eta^-_0) \, , \nonumber \\
\varphi'_{\vec k}(\eta^+_0)
&=& \varphi'_{\vec k}(\eta^-_0)
-  \Delta(\eta_H - \epsilon_H) {\cal H}_0 \, \varphi_{\vec k}(\eta_0)
\, ,
\label{bcsf}
\ea
where $\epsilon_H$ and $\eta_H$ are evaluated during inflation.
Here $\Delta(\eta_H - \epsilon_H) = (\eta_H^+ - \epsilon_H^+) -
( \eta_H^- - \epsilon_H^-) $ is the change of the differences of the
slow roll parameters after and before the transition. We note that
away from the transition, by the form of the action
(\ref{action}), we always have $\pi_{\vec k} =
\varphi'_{\vec k}$. Thus we can rewrite the jump conditions
(\ref{bcsf}) as the matching conditions for the fields and the
momenta at the transition:
\ba
\varphi_{\vec k}(\eta^+_0) &=&
\varphi_{\vec k}(\eta^-_0) \, , \nonumber \\
\pi_{\vec k}(\eta^+_0) &=& \pi_{\vec k}(\eta^-_0)
-  \Delta(\eta_H - \epsilon_H) {\cal H}_0 \,
\varphi_{\vec k}(\eta_0)
\, .
\label{bcs}
\ea
This makes the effect of the sudden transition very clear: it
enforces a canonical transformation on the variables
$\varphi_{\vec k},\pi_{\vec k}$ describing the inflaton dynamics
after the evolution has begun.
This in turn induces a Bogoliubov transformation between the
annihilation and creation operators. The Bogoliubov transformation
is proportional to the change in $\eta_H - \epsilon_H$ during the
transition. If inflation was short, such a change arises because
of the very nature of the slow roll regime. Namely, because during
inflation the curvature of the inflaton potential is small
compared to the Hubble scale, $\partial_\phi^2 V \le H^2$, and the
inflaton rolls slowly, if $\epsilon_H = \frac{\dot
\phi^2}{2m^2_{P} H^2} < 1, \eta_H = - \frac{\ddot \phi}{H\dot
\phi}<1$ during inflation, they will satisfy $\epsilon_H \ll 1,
\eta_H \ll 1$ before it. This is because $\dot \phi^2, \ddot \phi$
hardly changed at all, and $H^2 = \frac{\rho}{3m^2_P}$ was bigger
before. In the case of long inflation such a change can arise from
bumps in the inflaton potential or enhanced interactions with
matter sector for special values of the inflaton. However we
bear in mind that $\epsilon_H$ should not have changed by too much
since that would have produced a large variation of the leading
order result, that would render the subleading corrections which
we are considering essentially irrelevant.

Using (\ref{ancreops}) and (\ref{bcs}) we can write the
Bogoliubov transformation of the annihilation and creation operators
induced by the transition. Denoting those just before the transition
by their argument $\eta_0^-$ and those after by $\eta_0^+$,
the Bogoliubov transformation is
\be
\pmatrix{ b(\vec k, \eta_0^-) \cr
b^\dagger(-\vec k, \eta_0^-) \cr}
=
\pmatrix{ 1 + i \Delta (\eta_H - \epsilon_H) \frac{{\cal H}_0}{2k} &
i \Delta (\eta_H - \epsilon_H) \frac{{\cal H}_0}{2k} \cr
-i \Delta (\eta_H - \epsilon_H) \frac{{\cal H}_0}{2k} &
1 - i \Delta (\eta_H - \epsilon_H) \frac{{\cal H}_0}{2k} \cr}
\pmatrix{ b(\vec k, \eta_0^+) \cr
b^\dagger(-\vec k, \eta_0^+) \cr} \, .
\label{bogol}
\ee
We can now rewrite our initial inflaton state (\ref{defvac}) as
\be
b(\vec k, \eta_0^+) | \tilde 0 \rangle
= - i \Delta \bigl(\eta_H - \epsilon_H) \, \frac{{\cal H}_0}{2k} \,
b^\dagger(-\vec k, \eta_0^+) | \tilde 0 \rangle \, .
\label{state}
\ee
retaining only the terms of the order ${\cal
O}\bigl(\Delta(\eta_H - \epsilon_H)\frac{{\cal H}_0}{k},
\frac{1}{k^2\eta^2_0}\bigr)$, in accordance with our
approximations. We stress the key properties of this state. The
state (\ref{state}) is a direct and inevitable consequence of
evolution. It contains the contributions both from non-adiabatic
effects during the transition to the last stage of inflation, and
from the adiabatic dynamics during it. It is different from the
thermal vacuum, albeit by terms which vanish in the either of the
limits $\epsilon_H \rightarrow 0$, $\eta_0 \rightarrow - \infty$
and $ k \rightarrow \infty$. This means that in the limit of pure
de Sitter space and also at very short distances the quantum
correlations in $| \tilde 0 \rangle$ rapidly disappear. Hence the
theory defined by (\ref{bogol}) and (\ref{state}) is consistent
with decoupling. Although this comes from the suppressions by
$1/k^2 = - 1/\vec \nabla^2$, which resembles a non-local term, it
is automatically induced by the backreaction, and is perfectly
well behaved at short distances, where the theory may be cut off
in the usual way in order to regulate its UV behavior.

We can now compute the imprints of $ | \tilde 0 \rangle$ on the
inflationary fluctuations. Using $\pi_{\vec k} = \varphi'_{\vec
k}$, $\varphi_{\vec k}(\eta) = u_k(\eta) b(\vec k) + u^*_k(\eta)
b^\dagger(-\vec k)$ and the eigenmodes (\ref{eigensols}) we can
solve for the evolution of the Heisenberg operators $b(\vec k,
\eta), b^\dagger(\vec k, \eta)$ from the transition onwards. In
terms of the Schr\"odinger operators defined in
(\ref{scalarinflq}), they are
\ba
b(\vec k, \eta) &=& f_k(\eta) \, b(\vec k) +
g_k(\eta) \, b^\dagger(-\vec k)  \, , \nonumber \\
b^\dagger(-\vec k, \eta) &=&
f^*_k(\eta) \, b^\dagger(-\vec k)  +
g^*_k(\eta) \, b(\vec k) \, , \nonumber \\
f_k(\eta) &=& \sqrt{\frac{k}{2}} \, u_k(\eta)
+ \frac{i}{\sqrt{2k}} \, u'_k(\eta) \, , \nonumber \\
g_k(\eta) &=& \sqrt{\frac{k}{2}} \, u^*_k(\eta)
+ \frac{i}{\sqrt{2k}} \, {u^*_k}'(\eta) \, .
\label{solgen}
\ea
The functions $f_k, g_k$ obey $f^*_k f_k - g^*_k g_k = 1$ by
virtue of the Wronskian relation of the eigenmodes $u_k {u^*_k}' -
u_k' u^*_k = i$, and so (\ref{solgen}) in fact is the
evolution-induced adiabatic Bogoliubov rotation between $b(\vec
k), b^\dagger(\vec k)$ and $b(\vec k, \eta), b^\dagger(\vec k,
\eta)$. Using this general form of the solutions, it is
straightforward to obtain the evolution of the Heisenberg
operators from the time $\eta_0^+$ to $\eta$. One finds (dropping
the superscript $``+"$ on $\eta_0$ for notational simplicity)
\ba
b(\vec k, \eta) &=& U_k(\eta,\eta_0) \, b(\vec k, \eta_0) +
V_k(\eta, \eta_0) \, b^\dagger(-\vec k, \eta_0) \, , \nonumber \\
b^\dagger(-\vec k, \eta) &=&
U_k^*(\eta,\eta_0) \, b^\dagger(-\vec k, \eta_0) +
V_k^*(\eta, \eta_0) \, b(\vec k, \eta_0) \, ,
\label{solheis}
\ea
where
\ba
U_k(\eta, \eta_0) &=& f_k(\eta) f^*_k(\eta_0)
- g_k(\eta) g^*_k(\eta_0) \, ,
\nonumber \\
V_k(\eta, \eta_0) &=& g_k(\eta) f_k(\eta_0)
- f_k(\eta) g_k(\eta_0) \, .
\label{solall}
\ea
These functions satisfy
$U^*_k(\eta,\eta_0) U_k(\eta,\eta_0)
- V^*_k(\eta,\eta_0) V_k(\eta,\eta_0) = 1$
because $f^*_k f_k - g^*_k g_k = 1$,
and so they indeed also
comprise a time-dependent Bogoliubov transformation.
Now using (\ref{solheis})
we can finally write down the solution
for the field modes $\varphi_k(\eta) =
\frac{1}{\sqrt{2k}} \bigl(b_2(\vec k, \eta)
+ b_2^\dagger(-\vec k, \eta) \bigr)$
in terms of the Heisenberg operators
$b_2(\vec k, \eta_0), b_2^\dagger(\vec k, \eta_0)$:
\be
\varphi_{\vec k}(\eta) = \frac{1}{\sqrt{2k}} \Bigl( U_k(\eta,\eta_0)
+ V^*_k(\eta,\eta_0) \Bigr) \, b_2(\vec k, \eta_0) +
\frac{1}{\sqrt{2k}} \Bigl( U^*_k(\eta,\eta_0)
+ V_k(\eta,\eta_0) \Bigr) \, b^\dagger_2(-\vec k, \eta_0)   \,  .
\label{fieldsoln}
\ee
It is now straightforward albeit tedious to compute
the 2-point function of the operator (\ref{fieldsoln})
in the initial state $| \tilde 0 \rangle$ defined in
(\ref{state}). Expanding the initial state to the first
order in $\Delta(\eta_H-\epsilon_H)$ as given in (\ref{state}),
the result is
\ba
\langle \tilde 0 | \frac{\varphi_{\vec k}(\eta)}{a}
\frac{\varphi^\dagger_{\vec q}(\eta))}{a}
| \tilde 0 \rangle &=& \frac{1}{2k a^2} \,
\delta^{(3)}(\vec k - \vec q) \,
\Bigl\{ |U_k(\eta,\eta_0) + V^*_k(\eta,\eta_0)|^2  \, \nonumber \\
&+& {{\cal A}_k} \,
\Bigl(U_k(\eta,\eta_0) + V^*_k(\eta,\eta_0) \Bigr)^2
+ {{\cal A}^*_k} \,
\Bigl(U^*_k(\eta,\eta_0) + V_k(\eta,\eta_0) \Bigr)^2
\, \Bigr\}
 \, ,
\label{twopt}
\ea
where
\be
{\cal A}_k = - i \Delta \bigl(\eta_H - \epsilon_H)
\, \frac{{\cal H}_0}{2k} \, .
\label{biga}
\ee

These expressions however still contain higher powers of the slow
roll parameters and $(k\eta_0)^{-1}$ than is allowed by our
approximations. Therefore we need to organize the result
(\ref{twopt}) as a consistent perturbative expansion. Our
organizing principle is to view the result (\ref{twopt}) as the
standard 2-point function of the inflaton in the thermal vacuum,
plus small corrections coming from slow roll corrections and from
the initial correlations encoded in the definition of the initial
state of inflation (\ref{state}). This is a direct consequence of
our assumption that the inflaton state in which the fluctuations
are produced was the adiabatic vacuum just before the transition.
The symmetries of the approximate de Sitter space used to define
this vacuum then guarantee that the effects of the transition can
be organized as a perturbation series. The small dimensionless
numbers which characterize the corrections are: the slow roll
parameters $\epsilon_H$, $\eta_H$ alone, which account for the
fact that the apparent horizon is slowly growing during inflation,
the even powers of ${\cal H}_0/k = -\frac{1}{k \eta_0}$
controlling the adiabatic evolution and $\Delta (\eta_H -
\epsilon_H) \frac{{\cal H}_0}{k}$ controlling the magnitude of the
initial quantum correlations in $| \tilde 0 \rangle$ in
(\ref{state}). Because we are interested here only in comparing
these effects in the leading order, we will ignore the
interference between them. By our
choice of the inflaton state, which implicitly rests on the
validity of the slow roll approximation, these terms will be
subleading. If the slow roll
conditions are strongly violated, the interference terms may
become larger, but the leading order effects will be even more
important. Having assumed the validity of perturbation theory,
we can ignore this regime altogether.

To get the slow roll corrections on top of the thermal vacuum
result accounted for, we can take the limit of the 2-point
function (\ref{twopt}) where the state (\ref{state}) reduces to
the thermal vacuum. This amounts to taking ${\cal O}(\epsilon_H)$
terms in (\ref{twopt}) to zero, and taking $f_k(\eta_0)
\rightarrow 1$, $g_k(\eta_0) \rightarrow 0$. Using (\ref{solgen})
and (\ref{solall}) it is easy to verify that in this limit
$U_k(\eta, \eta_0) + V^*_k(\eta, \eta_0) \rightarrow \sqrt{2k}
u_k(\eta)$. Hence the 2-point function indeed reduces to the
thermal vacuum result,
\be
\langle \tilde 0 | \frac{\varphi_{\vec k}(\eta)}{a}
\frac{\varphi^\dagger_{\vec q}(\eta))}{a} | \tilde 0
\rangle \rightarrow \langle 0 | \frac{\varphi_{\vec k}(\eta)}{a}
\frac{\varphi^\dagger_{\vec q}(\eta))}{a} | 0 \rangle =
\frac{|u_k(\eta)|^2}{a^2} \, \delta^{(3)}(\vec k - \vec q)  \,  .
\label{twopttherm}
\ee
The slow roll corrections to the leading
order have been computed in \cite{lyste} by expanding the Hankel
functions (\ref{eigensols}) to the leading order in the slow roll
parameters at horizon crossing. We can simply take their result,
which with our conventions is
\be
\langle  0 | \frac{\varphi_{\vec k}(\eta)}{a}
\frac{\varphi^\dagger_{\vec q}(\eta))}{a} | 0 \rangle
= \frac{2 \pi^2}{k^3} \Bigl( \frac{H}{2\pi} \Bigr)^2 \, \Bigl( 1 +
2 \bigl(2 - \ln 2 - \gamma_{em} \bigr) (2 \epsilon_H - \eta_H ) -
2 \epsilon_H \Bigr) \, \delta^{(3)}(\vec k - \vec q) \, ,
\label{thermal}
\ee
where $\gamma_{em}= 0.5772 \ldots $ is the
Euler-Mascheroni constant.

Now we turn to the effects arising from the deviation of the
state (\ref{state}) from the thermal vacuum. From the discussion above,
it is clear that they are encased in the factor
\ba
{\cal F} &=& \frac{1}{2k |u_k(\eta)|^2} \,
\Bigl\{ |U_k(\eta,\eta_0) + V^*_k(\eta,\eta_0)|^2  \, \nonumber \\
&+& {{\cal A}_k} \,
\Bigl(U_k(\eta,\eta_0) + V^*_k(\eta,\eta_0) \Bigr)^2
+ {{\cal A}^*_k} \,
\Bigl(U^*_k(\eta,\eta_0) + V_k(\eta,\eta_0) \Bigr)^2
\, \Bigr\}
 \, ,
\label{factor}
\ea
which should be evaluated in the limit
$\eta \rightarrow 0$ using the relativistic
modes (\ref{modes}) in the
definitions of $f_k$, $g_k$, $U_k$ and $V_k$ above.  Hence
\ba
f_k(\eta) &=& \Bigl(1 - \frac{i}{k\eta}
- \frac{1}{2k^2\eta^2}) \, e^{-ik\eta}
\, , \nonumber \\
g_k(\eta) &=& \frac{1}{2k^2\eta^2}  \, \, e^{ik\eta}
\label{relfs}
\ea
To simplify the calculation,
note that $\frac{1}{\eta_0} = -{\cal H}_0$.
After the eviction from the horizon,
$u_k \rightarrow  - \frac{i}{\sqrt{2k}k\eta} e^{-ik \eta}$.
Thus, $f_k + g_k^* = \sqrt{2k} u_k \rightarrow
 - \frac{i}{k\eta} e^{-ik \eta}$. Therefore
\be
\Bigl( U_k(\eta, \eta_0) + V^*_k(\eta, \eta_0) \Bigr)_{|k\eta| \ll 1}
= -\frac{i}{k\eta} \, \Bigl( \bigl(1 - i\frac{{\cal H}_0}{k}
- \frac{{\cal H}_0^2}{2k^2} \bigr) e^{-ik/{\cal H}_0}
+ \frac{{\cal H}_0^2}{2k^2} e^{ik/{\cal H}_0} \Bigr)
\label{relvs}
\ee
Substituting this in ${\cal F}$ in (\ref{factor}), we find
(keeping only the terms up to
${\cal O}\bigl(\Delta(\eta_H - \epsilon_H) \frac{{\cal H}_0}{k},
\frac{{\cal H}^2_0}{k^2}\bigr)$),
\be
{\cal F} = 1 + \frac{{\cal H}_0^2}{k^2} \,
\cos(\frac{2k}{{\cal H}_0}) +
\Delta \bigl(\eta_H - \epsilon_H \bigr) \,
\frac{{\cal H}_0}{k} \, \sin(\frac{2k}{{\cal H}_0}) \, .
\label{factorhk}
\ee
In this expression, $\frac{{\cal H}_0}{k} = \frac{H}{p} \la 1$,
where we have defined the physical momentum of the mode $p =
\frac{k}{a_0}$ at the moment of the transition to the last $\sim 60$ e-folds
of inflation. This is simply the statement that the fluctuations
are produced inside the inflating Hubble patch, meaning that their
wavelength is originally inside the apparent horizon.
We can rewrite the form factor ${\cal F}$
in terms of these variables as
\be
{\cal F} = 1 + \frac{H^2}{p^2} \,
\cos(\frac{2p}{H}) +
\Delta \bigl(\eta_H - \epsilon_H \bigr) \,
\frac{H}{p} \, \sin(\frac{2p}{H}) \, .
\label{factorfin}
\ee
In this parameterization, the unity corresponds to the thermal
vacuum result, the second term (proportional to ${\cal O}\Bigl((H/p)^2\Bigr)$)
to the adiabatic evolution of the vacuum, and the third
term (proportional to ${\cal O}\Bigl(\Delta(\eta_H - \epsilon_H)(H/p)\Bigr)$)
to the effects of the quantum correlations encoded in
the state $| \tilde 0 \rangle$ (\ref{state}) by the
transition to the final stage of inflation.

To see the total effect on the density fluctuations, we fold
(\ref{factorfin}) with the slow roll-corrected 2-point function in
the thermal vacuum. Again expanding to the linear order in all
corrections, we obtain the full 2-point function at horizon
crossing:
\be
\langle \tilde 0 | \frac{\varphi_{\vec k}(\eta)}{a}
\frac{\varphi^\dagger_{\vec q}(\eta))}{a}
| \tilde 0 \rangle  = \frac{2 \pi^2}{k^3}
\Bigl( \frac{H}{2\pi} \Bigr)^2 \,
\Bigl( 1 + {\cal D}(p, H, \epsilon_H, \eta_H) \Bigr)
\, \delta^{(3)}(\vec k - \vec q) \, ,
\label{twoptcorr}
\ee
where
\be
{\cal D}(p, H, \epsilon_H, \eta_H) =
2 \bigl(2 - \ln 2 - \gamma_{em} \bigr)
(2 \epsilon_H - \eta_H )
- 2 \epsilon_H + \frac{H^2}{p^2} \,
\cos(\frac{2p}{H}) + \Delta
\bigl(\eta_H - \epsilon_H \bigr) \,
\frac{H}{p} \, \sin(\frac{2p}{H}) \, .
\label{twoptfin}
\ee
Substituting this into the formula for the power
spectrum (\ref{power}) yields for modes with $k\eta_0 = p/H > 1$,
which were expelled out of the horizon,
\be
{\cal P}(k) = \Bigl(\frac{H}{\dot \phi}\Bigr)^2 \,
\Bigl( \frac{H}{2\pi} \Bigr)^2 \,
\Bigl( 1 + {\cal D}(p, H, \epsilon_H, \eta_H) \Bigr) \, ,
\label{powerfin}
\ee
which is our main result.

This shows that if there is a sudden change in $\eta_H -
\epsilon_H$, either because inflation is short, lasting not much
more than the minimum 60 e-folds, or because $\eta_H$ jumps $\sim
60$ e-folds before the end of long inflation while $\epsilon_H$
remains roughly constant, which would keep the leading order
result unchanged and confine the interesting effects in the
corrections we derive, the imprints of this change in the
inflationary fluctuations could be at a level that may affect the
observations. The perturbations from the modes which are expelled
out of the horizon just after this transition may receive
important contributions from the quantum correlations inside the
inflating patch, which would be comparable with the slow roll and
adiabatic effects. As inflation progresses and shorter and shorter
wavelength modes are expelled out of the horizon, these
corrections will rapidly diminish below the observable level. That
is clear from (\ref{twoptfin}), (\ref{powerfin}) because the
contributions from these correlations are suppressed by $H/p$ as
the momentum increases. We interpret this as the quantum version
of the cosmic no-hair theorem: as inflation proceeds the quantum
state of the inflaton, out of which the fluctuations emerge, is
less and less different from the thermal vacuum. We stress again
that while the effect in (\ref{twoptfin}), (\ref{powerfin}) might
be vaguely reminiscent of the corrections claimed to arise in the
$\alpha$-vacua \cite{danielsson}, they really are completely
different. Our signal explicitly depends on the physical momentum
$p$, or the wavelength of the perturbations $\lambda = 1/p$,
rather than on some fixed trans-Planckian cutoff. Further, we find
that there is a suppression by the change of the difference of the
slow roll parameters $\Delta (\eta_H - \epsilon_H)$. Thus our
effects only appear in the long wavelength perturbations, and
rapidly vanish in the UV. In this way, our results are fully
consistent with the conventional lore of effective field theory,
because decoupling of the UV physics is guaranteed.

Nevertheless, the result (\ref{twoptfin}), (\ref{powerfin}) has
interesting implications for observational cosmology. At present,
the case for inflation is growing stronger as more data are
accumulated \cite{boomer,maxima,wmap}. However it is difficult to
use observations to place bounds on the duration of inflation, or
the properties of the inflationary potential. Our result may serve
as an additional probe of the inflationary dynamics. While
somewhat model-dependent, our result suggests that in the case of
either short inflation or longer inflation with a large $\Delta
(\eta_H - \epsilon_H)$ the density spectrum, and therefore the
CMB, may retain some information about the initial quantum
correlations at the instant when this stage began. There may be
models where such terms could be at the level of few percents, and
therefore observable. Because the effects in
(\ref{twoptfin}), (\ref{powerfin}) come with a distinct trigonometric
modulation at the largest scales, this might help in the search for
them. Note that at shorter scales, as the momentum increases, the
modulation essentially disappears: the statistical sampling of the
data tells us that we must average the  trigonometric functions over
several periods. This would render the modulation at short scales
impossible to detect, and therefore completely irrelevant.

In some models, the effects leading to (\ref{twoptfin}) and
(\ref{powerfin}) may even suppress power on large scales, reducing
the low $\ell$ multipoles in the CMB anisotropy. These multipoles
are sensitive to scales of the order of the horizon today and
larger, and so in short inflation they could be affected by the
superhorizon modes at the onset of inflation, obeying $k\eta_0 =
p/H < 1$. Although one does not have firm control over the
fluctuations on scales $p/H \ll 1$ because they would be strongly
affected by any initial inhomogeneities outside the inflating
patch, we can at least estimate how much power would be
transferred to these modes by inflationary dynamics if the initial
inhomogeneities were negligible. In this regime we can neglect
slow roll corrections and the
${\cal\,O}(\Delta(\eta_H-\epsilon_H))$ term encoding initial
quantum correlations. Then using the result for the power spectrum
including adiabatic corrections, valid on all scales, ${\cal
P}(k)=(H/2\pi)^2\,(H/\dot \phi)^2\,
\{1+H^2/2p^2+(H^2/p^2-H^4/2p^4)\cos(2p/H)-(H/p)^3\sin(2p/H)\}$, we
find the leading order power spectrum for the superhorizon modes
at the onset of inflation by taking the limit $p/H < 1$,
\begin{equation}
{\cal P}(k)={4 \over 9} \Bigl(\frac{H}{\dot \phi}\Bigr)^2 \,
\Bigl( \frac{H}{2\pi} \Bigr)^2 \Bigl(\frac{p}{H}\Bigr)^2 \,.
\label{powerlargescale}
\end{equation}
This shows that the deposit of power in superhorizon modes during
inflation is strongly suppressed, as expected. Thus for short
inflation, the reduced amplitude of the low $\ell$ multipoles in
the CMB anisotropy arises from combining (\ref{powerfin}) for
$p>H$ and (\ref{powerlargescale}) for $p<H$.

One may correctly warn that the trigonometric modulation present
in Eqs. (\ref{twoptfin}), (\ref{powerfin}) need not be an
unambiguous indication of the presence of short inflation or long
inflation with a jump in $\eta_H - \epsilon_H$. For example, one
may try to redefine the background of the theory by redefining the
inflationary potential $V(\phi) \rightarrow W(\phi)$ by solving
the differential equation
\be
\frac{\partial_\phi W}{W^{3/2}} =
\frac{\partial_\phi V}{V^{3/2}} \,
\Bigl( 1 + {\cal D}(p, H, \epsilon_H, \eta_H) \Bigr)^{-1/2} \, .
\label{potredef}
\ee
Thus our effects might be mimicked by a different potential, where
they would be confined in the leading order result. However if the
effects we are discussing are quantitatively significant, this
redefinition of the potential would not {\it remove} the question;
it would merely {\it change} it. One would still be forced to ask
``What produced such features in the inflaton potential $\sim 60$
e-folds before the end of inflation, which gave rise to such
signals?" regardless of the root cause of the signal itself. Hence
the presence of such effects would indicate interesting physics
either way! Their detection would be a win-win situation. On the
other hand, if no such effects are ever seen, it would be natural
to argue that inflation went on uninterrupted for significantly
more than the bare minimum of $\sim 60$ e-folds. This would bury
any information in the inflationary perturbations about the
initial state below the discernible level. However, in such an
instance one could plausibly argue that the curvature of the
spatial sections is very tiny, and therefore that the density of
dark energy plus dark matter is practically indistinguishable from
unity.

In summary, we have shown how quantum correlations in the quantum
state of the inflaton affect the density perturbations and the
CMB. Our calculations are in full agreement with the usual
effective field theory and decoupling, and are performed in the
controllable regime of perturbation theory, where the universe can
be treated as a weakly perturbed FRW cosmology. The effects of
these correlations are suppressed by a power of the momentum, and
vanish in the UV. We find that if inflation didn't last much
longer than the necessary minimum of 60 or so e-folds, or if
$\eta_H - \epsilon_H$ changed significantly at that time, the
initial correlations may yield observable imprints. Thus the
amplitude of the corrections from these quantum correlations may
be a sensitive probe of the inflationary dynamics at $\sim 60$
e-folds before the end. Viewing the issue of the inflationary
dynamics during the final stage as a purely observational matter,
we feel that the prospective searches for such effects would be a
worthy enterprise, since they could shed light on the darkness
from which our universe emerged.

\vskip.5cm

{\bf Acknowledgements}

We would like to thank A. Albrecht, V. Balasubramanian, C.P.
Burgess, M.B. Einhorn, R. Holman, W. Hu, L. Hui, M. Kleban, A.
Lawrence, M. Santos, Dj. Mini\' c, S. Shenker, D. Spergel and
especially J.D. Bjorken and L. Susskind for useful discussions.
N.K. is supported in part by a Research Innovation Award from the
Research Corporation. M.K. is supported in part by the NASA grant
NAG5-11098.

\end{document}